\documentstyle[12pt]{article}

\input{tcilatex}

\begin{document}

\title{Laplace Transform of Spherical Bessel Functions}
\author{ A. Ludu \\
{\normalsize {Department of Chemistry and Physics, }}\\
{\normalsize {Northwestern State University,} } {\normalsize {Natchitoches,
LA 71497} } \and  R. F. O'Connell \\
Department of Physics and Astronomy,\\
Louisiana State University, Baton Rouge, LA 70803-4001}
\date{}
\maketitle

\begin{abstract}
We provide a simple analytic formula in terms of elementary functions for
the Laplace transform ${\tilde{j}}_{l}(p)$ of the spherical Bessel function
than that appearing in the literature, and we show that any such integral
transform is a polynomial of order $l$ in the variable $p$ with constant
coefficients for the first $l-1$ powers, and with an inverse tangent
function of argument $1/p$ as the coefficient of the power $l$. We apply
this formula for the Laplace transform of the memory function related to the
Langevin equation in a one-dimensional Debye model.
\end{abstract}

 \vfill$^{1}$ E-mail address:
ludua@nsula.edu

$^2$ E-mail address: phrfoc@lsu.edu

\vfill \eject

\bigskip

\vskip1.5cm {\bf I. Introduction. The memory function in the Debye model. } %
\vskip1.5truecm

In recent years, there has been widespread interest in dissipative problems
arising in a variety of areas in physics, and we refer to [1] for a
mini-review. As it turns out, solutions of many of these problems are
encompassed by a generalization of Langevin's equation to encompass quantum,
memory and non-Markovian effects, as well as arbitrary temperature and the
presence of an external potential V(x). As in [2], we refer to this as the
generalized quantum Langevin equation (GLE):

\begin{equation}
m\ddot{x}+\int_{-\infty }^{t}dt^{\prime }\mu {(}t-t^{\prime })\dot{x}%
(t^{\prime })+V^{\prime }(x)=F(t)+f(t),
\end{equation}

where $V^{\prime }(x)=dV(x)/dx$ is the negative of the time-independent
external force and $\mu (t)$ is the so-called memory function. $F(t)$ is the
random (fluctuation or noise) force and $f(t)$ is a c-number external force
(due to a gravitational wave, for instance). In addition (keeping in mind
that measurements of $\Delta {x}$ generally involve a variety of readout
systems involving electrical measurements), it should be strongly emphasized
that ''-- the description is more general than the language --'' [2] in that 
$x(t)$ can be a generalized displacement operator (so that, for instance, $%
\Delta {x}$ could represent a voltage change). Furthermore, $\mu (t)$ and $%
F(t)$ are given in terms of the parameters of the heat bath only. Explicitly

\begin{equation}
\mu (t)=\sum_{j}m_{j}\omega _{j}^{2}\cos (\omega _{j}t)\theta (t),
\end{equation}

where $\theta {(t)}$ is the Heaviside step function. Also

\begin{equation}
F(t)=\sum_{j}m_{j}\omega _{j}^{2}q_{j}^{h}(t),
\end{equation}

where $q^{h}(t)$ denotes the general solution of the homogeneous equation
for the heat-bath oscillators (corresponding to no interaction). Thus, we
have all the tools necessary for the analysis of any heat bath. As
emphasized in Refs. [1,3] of primary interest for the calculation of
observable physical quantities are either the Fourier or Laplace transforms
of $\mu (t)$. In particular, the blackbody radiation heat bath was
investigated in detail in Refs. [2,3].

Here, we wish to consider the one-dimensional Debye model [4] because it
leads to a result for the memory function which involves spherical Bessel
functions.

In this model one takes all the heat bath oscillators to have the same mass
i.e. $m_{j}\equiv {m}$ for all $j$. Thus, taking the maximum allowed heat
bath frequency to be $\omega _{L}$ we see that Eq. (2) reduces to

\begin{equation}
\mu (t)=m\sum_{j}\omega _{j}^{2}\cos \omega _{j}t=m\int_{0}^{\omega
_{L}}d\omega {D}(\omega )\omega ^{2}\cos \omega {t}
\end{equation}

where [4]%
\begin{equation}
D(\omega )=\frac{L}{\pi {v}},
\end{equation}

is the density-of-states. Also, $v$ is the velocity of sound and $L$ is the
length of a one-dimensional line such that $\omega =vk$ where $\Delta
k=(2\pi /L)$ is the interval between the allowed values of the wave-vector $%
k $. It follows that%
\begin{equation}
\mu (t)=\frac{mL}{\pi v}\int_{0}^{\omega _{L}}d\omega \omega ^{2}\cos \omega
t=\frac{mL}{3\pi v}\omega _{L}^{3}\{j_{0}(\omega _{L}t)-2j_{2}(\omega
_{L}t)\}
\end{equation}

Thus, in order to calculate $\tilde{\mu}(p)$ we will turn to our general
result given in the next section and related to Laplace transform of
spherical Bessel functions. For example, for the above case, the following
expression for the Laplace is obtained%
\begin{equation}
\tilde{\mu}(p)=\frac{mL}{\pi v}\omega _{L}p-p^{2}\tan ^{-1}\left( \frac{%
\omega _{L}}{p}\right) ,.
\end{equation}%
from a more general formula. For a detailed discussion of this result we
refer to Ref.[5].

\bigskip

\bigskip \vskip1.5cm {\bf II. Analytic formula for the Laplace transform. } %
\vskip1.5truecm

In order to calculate eq.(7), as well as the Laplace transform of higher
order Bessel functions for other applications, we introduce in the following
an exact formula, much simpler than formulas found in literature, in terms
of trigonometric functions and polynomials.

The spherical Bessel functions are given by (Ref. 6, p. 965) 
\begin{equation}
j_{l}(t)=\left( {\frac{{\pi }}{{2t}}}\right) ^{\frac{1}{2}}J_{l+{\frac{1}{2}}%
}(t),
\end{equation}%
where $l$ is a positive integer and $J_{\nu }$ are the Bessel functions of
the first kind of real argument $t$. The Laplace transform, defined by 
\begin{equation}
L[j_{l}(t)]\equiv {\tilde{j}}_{l}(p)=\int_{0}^{\infty }j_{l}(t)e^{-pt}dt,
\end{equation}%
can be calculated by using the relation [Ref. 7, p. 182, II (9)] 
\begin{equation}
L[t^{\mu }J_{\nu }(t)]=\Gamma (\mu +\nu +1){\frac{{\ P_{\mu }^{-\nu }}\left( 
\frac{p}{\sqrt{p^{2}+1}}\right) }{(p^{2}+1)^{\frac{{\mu +1}}{2}}}},
\end{equation}%
where $P_{\mu }^{\nu }$ are the Legendre functions and we have to fulfill
the restrictions: $\mu +\nu >-1$ and $p>0$. Also, $\Gamma $ is the Gamma
function defined [6] as \bigskip $\Gamma (x)=\int_{0}^{\infty
}e^{-t}t^{x-1}dt$ for positive values of $x$.

\ \ In the case of the spherical Bessel functions we have $\nu =l+1/2$ and $%
\mu =-1/2$ and hence the restriction is $l>-1$ which is always fulfilled
since $l=0,1,\dots $. Hence 
\[
{\tilde{j}}_{l}(p)=\sqrt{\frac{{\pi }}{2}}\Gamma (l+1)\left( {\frac{\rho }{p}%
}\right) ^{\frac{{1}}{{2}}}P_{-{\frac{{1}}{2}}}^{-l-{\frac{{1}}{2}}}(\rho ) 
\]%
\begin{equation}
=\sqrt{\frac{\pi }{2}}{\frac{{\Gamma (l+1)}}{{\Gamma (l+3/2)}}}\left( \frac{%
\rho }{p}\right) ^{\frac{1}{2}}\left( \frac{{1-\rho }}{{1+\rho }}\right) ^{%
\frac{{l+{\frac{1}{2}}}}{2}}F\left( {\frac{1}{2}},\ \ {\frac{1}{2}};\ \ l+{%
\frac{3}{2}};\ \ {\frac{{1-\rho }}{2}}\right) ,
\end{equation}%
where $\rho =p/\sqrt{p^{2}+1}$ and $p>0$ so that $1>\rho >0$. Here $F\left(
\alpha ,\ \beta ;\ \gamma ;\text{ }x\right) $ is the Gauss hypergeometric
series defined by the formula [6]

\[
F\left( \alpha ,\ \beta ;\ \gamma ;\text{ }x\right) =\sum_{k=0}^{\infty }%
\frac{\alpha ^{(k)}\beta ^{(k)}}{\gamma ^{(k)}}\cdot \frac{z^{k}}{k!}, 
\]%
where $\alpha ^{(k)}=\alpha (\alpha +1)...(\alpha +k-1);\alpha ^{(0)}=1$.

The formula provided by eq.(11) is somehow difficult to use in the cases $%
l>2 $, and requires tedious calculations and further manipulations such as
integration or other transforms. However, in the case $l=0$, eq.(11) reduces
to 
\begin{equation}
{\tilde{j}}_{0}(p)=\sqrt{2}\left( \frac{{\rho }}{p}\right) ^{\frac{1}{2}%
}\left( \frac{{1+\rho }}{{1-\rho }}\right) ^{-{\frac{1}{4}}}F\left( {\frac{1%
}{2}},\ \ {\frac{1}{2}};\ \ {\frac{3}{2}};\ \ {\frac{{1-\rho }}{2}}\right) .
\end{equation}%
Since the Gauss hypergeometric series in eq.(12) can be writen in the form
(Ref. 6, p. 1041 II 13) 
\begin{equation}
F\left( {\frac{1}{2}},\ \ {\frac{1}{2}};\ \ {\frac{3}{2}};\ \ \rho \right) ={%
\frac{\sin {^{-1}\sqrt{\rho }}}{\sqrt{\rho }}},
\end{equation}%
we find for the Laplace transform of $j_{0}(t)=\sin t/t$ the usual
expression (Ref. 7, p. 152) 
\begin{equation}
{\tilde{j}}_{0}=\tan ^{-1}{\frac{1}{p}}.
\end{equation}%
Another possibility for expressing the Legendre function for the $l=0$ case
in a simpler way is to use the result ( Ref. 6, p. 1008) 
\begin{equation}
P_{-{\frac{1}{2}}}^{-{\frac{1}{2}}}(\rho )=\lim_{q\rightarrow 0}P_{q-{\frac{1%
}{2}}}^{-{\frac{1}{2}}}(\rho )=\cos ^{-1}(\rho )\sqrt{\frac{{2}}{{\pi \sqrt{%
1-\rho ^{2}}}}}.
\end{equation}%
Finally, one can use a direct approach, based on the fact that the Fourier
representation of the spherical Bessel functions are definite integrals over
Legendre poynomials 
\[
j_{l}(t)={\frac{1}{2}}(-i)^{l}\int_{-1}^{1}e^{ilx}P_{l}(x)dx. 
\]%
The following step is to perform the integration for the Laplace transform,
obtaining the relation ${\tilde{j}}_{l}(p)=i^{l+1}Q_{l}(ip)$, where $%
Q_{l}(p) $ are the Legendre functions of second type [6-8].

However, none of these approaches can provide an analytic expression in
terms of elementary functions for the $l\geq 2$ case. Thus, we are motivated
to develop a new approach based on recursion relations. By differentiating
two times the second order spherical Bessel function 
\begin{equation}
j_{2}=\left( \frac{3}{t^{2}}-1\right) {\frac{{\sin t}}{{t}}}-{\frac{2{\cos t}%
}{{\ t^{2}}}},
\end{equation}%
and by using recurrently the formula for the Laplace transform of the
derivative of a function $f(t)$ (Ref. 7, p. 129) 
\begin{equation}
{\tilde{f^{\prime }}}(p)=p{\tilde{f}}(p)-f(0),
\end{equation}%
where the prime denotes the derivative with respect to $p$, we can calculate
the Laplace transform of $j_{2}.$That is we choose for $f(t)=\sin \alpha t/t$
which has its Laplace transform in tables [6,7]. We differentiate $f(t)$
once and we calculate the Laplace transform of $f^{\prime }(t)$ by using
Eq.(17). Then, we differentiate one more time and find again the Laplace
transform of $f^{\prime \prime }(x)$ by formula Eq.(17). Since we can
express $j_{2}$ only in terms of the functions $f(t)$ and $f^{^{\prime
\prime }}(x)$ we can express its Laplace transform in terms of the Laplace
transforms of $f(t),f^{\prime \prime }(t)$. We finally obtain 
\begin{equation}
{\tilde{j}}_{2}=\left( {\frac{{3p^{2}}}{{2}}}+{\frac{{1}}{{2}}}\right) \tan
^{-1}{\frac{{1}}{{p}}}-{\frac{{3p}}{{2}}}.
\end{equation}

This procedure gives us a hint for constructing a recursion formula for the
Laplace transform of the spherical Bessel functions of any order. This will
provide a much simpler form than that one provided by eq.(11). This approach
needs only the derivative recursion formula, eq.(17), and the transform of
the zero order function. By starting from the recursion formulas for the
spherical Bessel functions (Ref. 6, p. 967) 
\[
j_{l+1}(\rho )=-j_{l}^{\prime }(\rho )+{\frac{l}{{\rho }}}j_{l}(\rho ) 
\]%
\begin{equation}
j_{l+1}(\rho )={\frac{{2l+1}}{{\rho }}}j_{l}(\rho )-j_{l-1}(\rho ),\ \ l\geq
1,
\end{equation}%
and by writing them in a convenient form 
\begin{equation}
j_{l+1}=-{\frac{{2l+1}}{{l+1}}}j_{l}^{\prime }+{\frac{l}{{l+1}}}j_{l-1},
\end{equation}%
we find that any spherical Bessel function can be expressed as a sum of the
derivatives of different orders of $j_{0}$ 
\begin{equation}
j_{l}(\rho )=\sum_{k=0}^{2k\leq l}C_{k}^{(l)}{\frac{{d^{l-2k}}}{{d\rho
^{l-2k}}}}j_{0}(\rho ),
\end{equation}%
where $C_{k}^{(l)}$ are coefficients depending on the labels $l$ and $k$. In
order to calculate these coefficients we use again the recursion relation
eq.(20) and the formula in eq.(21) and we obtain the equation 
\begin{equation}
\sum_{k=0}^{2k\leq l}C_{k}^{(l)}j_{0}^{(l-2k+1)}={\frac{{l}}{{2l+1}}}%
\sum_{k=0}^{2k\leq l-1}C_{k}^{(l-1)}j_{0}^{(l-2k-1)}-{\frac{{l+1}}{{2l+1}}}%
\sum_{k=0}^{2k\leq l+1}C_{k}^{(l+1)}j_{0}^{(l-2k+1)},
\end{equation}%
where the superscript on the Bessel function indicates the order of
differentiation. Since the Wronskian form of any two derivatives of $j_{0}$
is not identical to zero, all these derivatives are functional independent
so we can determine the numbers $C_{k}^{(l)}$ in eq.(22) by indentifying the
coefficients of the same order of derivative of the Bessel function.
Consequently we find the binary recursion relation 
\begin{equation}
C_{0}^{l+1}=-{\frac{{2l+1}}{{l+1}}}C_{0}^{(l)},
\end{equation}%
which provide all $k=0$ coefficients 
\begin{equation}
C_{0}^{(l+1)}=(-1)^{l+1}{\frac{{(2l+1)!!}}{{(l+1)!}}},\ \ \ l=0,\ 1,...\ \ 
\text{and}\ \ C_{0}^{(0)}=0.
\end{equation}%
and a ternary recursion relation 
\begin{equation}
C_{k}^{(l+1)}=-{\frac{{2l+1}}{{l+1}}}C_{k}^{(l)}+{\frac{l}{{l+1}}}%
C_{k-1}^{(l-1)},\ \ \ l=1,\ 2,...,\ \ k\leq l,
\end{equation}%
which provides the other coefficients. Also, if we are given $C_{1}^{(2)}$
(which can be easily obtained by giving particular values in eq.(22), that
is $C_{1}^{(2)}=1/2$) we can generate all the coefficients by using the
ternary relation, eq.(25). By using this procedure it is easy to identify
all coefficients in the formula eq.(21). For example we have 
\[
j_{1}=-j_{0}^{(1)},\ \ \ j_{2}={\frac{3}{2}}j_{0}^{(2)}+{\frac{1}{2}}j_{0},\
\ \ j_{3}=-{\frac{{5}}{2}}j_{0}^{(3)}-{\frac{3}{2}}j_{0}^{(1)}, 
\]%
\begin{equation}
j_{4}={\frac{{35}}{8}}j_{0}^{(4)}+{\frac{{15}}{4}}j_{0}^{(2)}+{\frac{3}{8}}%
j_{0},\ \ \ \text{etc.}
\end{equation}%
The next step is to calculate the Laplace transform by using eqs.(17) and
(21). Since 
\begin{equation}
j_{0}^{(2k)}(0)={\frac{{(-1)^{k}}}{{2k+1}}},\ \ j_{0}^{(2k+1)}(0)=0,
\end{equation}%
and from eqs.(14) and (17), we have 
\begin{equation}
L[j_{0}^{(n)}]=p^{n}\tan ^{-1}{\frac{1}{{p}}}%
-\sum_{m=0}^{n-1}p^{n-m-1}j_{0}^{(m)}(0),
\end{equation}%
we finally obtain the desired formula 
\begin{equation}
{\tilde{j}}_{l}(p)=p^{l}\left[ \tan ^{-1}{\frac{1}{p}}\cdot
\sum_{k=0}^{2k\leq l}C_{k}^{(l)}p^{-2k}-\sum_{k=0}^{2k\leq
l}C_{k}^{(l)}\sum_{m=0}^{[{\frac{{l-1}}{2}}]-k}{\frac{{(-1)^{m}}}{{2m+1}}}%
p^{-2k-2m-1}\right] ,
\end{equation}%
valid for $l\geq 1$, where $[\ \cdot \ ]$ stands for the integer part. The
Laplace transform is hence given in terms of elementary functions, by a
polynomial $Q$ of order $l-1$ in the variable $p$, with constant
coefficients (all coefficients determined by the recursion relations
eqs.(24) and (25)) plus another polynomial $P$ of order $l$ multiplied by
the inverse tangent function of $1/p$ 
\begin{equation}
{\tilde{j}}_{l}=P_{l}(p)\tan ^{-1}(1/p)+Q_{l-1}(p).
\end{equation}
Finally, we compare the above algorithm, based on recurrsion relations
Eqs.(24, 25) and Eqs. (29, 30), with direct numerical algorithms used in the
corresponding built-in functions in different symbolic programs, like for
instance {\it Mathematica. }All functions involved above have their
equivalent built-in correspondent in {\it Mathematica-4.0 (}used on a Power
Macintosh 8500 computer), that is {\it Gamma[z], BesselJ[n,z],
LegendreP[n,z], LegendreQ[n,m,z], }etc. While comparing the CPU time elapsed
for both procedures we find out that our recurrsion approach is faster. That
is the relative difference between the two CPU intervals of time increases
as $e^{3.2l}$ (numerical with respect to analytic), where $l$ is the order
of the spherical Bessel function.

\bigskip

\vskip1.5cm {\bf III. Conclusions} \vskip1.5truecm

\bigskip

By using a generalization of the quantum Langevin's equation in terms of
memory function one can obtain exact solutions for dissipative problems
arising in many areas in physics. In terms of a one-dimensional Debye model
the memory function is expressed as a combination of spherical Bessel
functions. However, of primary interest for the claculation of physical
quantities are the laplace (or Fourier) transform of the memory function.

Also, the Schr\"{o}dinger equation for a free particle in polar coordinates
leads, for each value of the positive integer $l$ of the orbital angular
momentum, to a radial equation which results in the generic equation for
spherical Bessel functions. In electrodynamics, too, the spherical Bessel
functions are related to solutions of the field equations in the stationary
or quasi-stationary regime in cylindrical geometry. And moreover, Bessel
functions are involved in the Helmholtz equation in cylindrical coordinates.
In all such applications one needs the integral transforms (especially the
Fourier or Laplace transforms) of such solutions. Constructing an exact
analytic formula for such applications is the object of our studyy.
Consequently, the present letter is not addressed to experts in special
functions, but rather to physicists who just need to apply such formulas in
their research, in the simplest possible form.

We have obtained a simpler exact formula for the Laplace transform of the
Spherical Bessel functions of any order, in terms of polynomial and
trigonometric functions, that is in terms of elementary functions. Among
other representations of such functions, this expression is much simpler,
hence more useful for potential applications. The speed of calculation is
one of foremost importance for applications similar with the one presented
above, especially in more than one dimension. We compared this formula with
some numerical evaluation in terms of precision and suitability for faster
implementation. Our comparison shows that the introduced formula is easy to
be used and it introduces less computational complexity than conventional
numerical techniques.

\bigskip

\vskip 2cm {\bf Aknowledgments} \vskip 1cm The authors are thankful to P.
Abbott, The University of Western Australia, for his valuable comments.

\end{document}